# A simple technique for the automation of bubble size measurements

*T. Gaillard[1], C. Honorez[1], M. Jumeau[1], F. Elias[2,3], W. Drenckhan[1]*

1. Laboratoire de Physique des Solides, UMR8502 – Université Paris Sud, Orsay, France
2. Matière et Système Complexes, Université Paris 7– UMR7057, France
3. Université Pierre et Marie Curie, Paris, France

**Abstract**

An increasing number of research topics and applications ask for a precise measurement of the size distribution of small bubbles in a liquid - and hence for reliable and automated image analysis. However, due to the strong mismatch between the refractive index of a liquid and a gas, bubbles deform strongly the path of light rays, rendering automated bubble size analysis a challenging task. We show here how this challenge can be met using the fact that bubbles act like inverted, spherical lenses with a curvature which is the inverse of the bubble radius. The imaging properties of each bubble can then be used to accurately determine the radius of the bubble upon imaging an object which can be filtered easily by a computer. When bubbles are large enough to be deformed under the influence of gravity, it is more appropriate to measure their size after squeezing them between two narrowly spaced glass plates. We therefore show here, how the analysis can be extended to this case; and how both approaches can be combined to measure the size distributions of strongly polydisperse foams containing simultaneously small (several 10s of micrometres) and large bubbles (several 100s of micrometres).

# 1 Introduction

Bubbles of sub-millimetric dimensions play an important role in an increasing number of applications such as foam flotation [1], foam fractionation [2] or bubble column reactors [3]. They also play an important role in many geophysical problems [4]. The bubble sizes as well as the size distributions play an important role in determining the final properties of the fluid/bubble mixture and therefore need to be known with sufficient precision. For this purpose, numerous techniques have been developed in the past, including liquid scattering [5], acoustic methods, or laser scanning techniques (see introductory review in [6]). These techniques are efficient, but generally suffer from a lack in precision when reliable information about the size distribution is required. Due to their simplicity and low cost, classic photographic techniques making use of white light are the preferred tool for precise bubble size measurements. However, due to the strong mismatch of the refractive indices between the gas and the liquid, and the curved surfaces of the bubbles, they interfere strongly with the light paths and therefore render quantitative imaging a challenging task [7]. This is even more the case when automated image analysis is required with the desire to obtain reasonably reliable statistical measures on averages and distributions.

In order to overcome this challenge, we propose here two simple approaches. The first (Section 0) considers freely floating bubbles which are small enough not to be deformed by gravity (for example several 10s of micrometres in water). To measure their size, a simple approach can be used which exploits the fact that a bubble in a liquid acts like an inverse, spherical lens which creates a virtual image which can be photographed by a camera. The relationship between the size of the final image and the imaged object is directly related to the curvature of the bubble and hence its radius. By exploiting the curvature of the bubble, the image treatment does not need to look for the boundary of the bubble which tends to be ill-defined in most lighting conditions [7]. The geometry of the imaged object needs to be defined in a manner that it can be picked up easily by a computer program amongst other objects (reflections, space between bubbles) in the overall image. The simplest shape is a circle – which is what we use here in order to illustrate the approach. The second approach (Section 4) deals with bubbles which are large enough to be deformed under gravity (several 100s of micrometres). In this case it is more adapted to squeeze them between two narrowly spaced glass plates, creating so-called bubble "pancakes". Using surface evolver simulations [8] (http://www.susqu.edu/brakke/evolver/evolver.html) of such squeezed bubbles, we show how their undeformed sizes can be determined from simple photographs using the same experimental set-up as for the small bubbles. I.e. both, small and large bubbles of a strongly polydisperse bubble mixture or foam can be measured simultaneously and in a fully automated manner.

## 2   Imaging properties of a bubble

A spherical bubble of radius $R_B$ surrounded by a liquid acts like an inverse, spherical lens, i.e. like a collector lens. As indicated in Figure 1a (scheme not drawn to scale), an object of size $R_L$ creates a smaller, virtual image of size $R_C$ and of the same orientation between the lens and the focal point "$f$" of the bubble lens. Figure 1b illustrates this effect by showing the example of the letter "R" on a bright, square-shaped background, imaged through three air bubbles in water (left image). On the right of Figure 1b, the image of the same letter "R" is shown through three water drops in air. These images are created using the ray tracing software Studio Max (http://www.autodesk.com/, as explained in detail in [7]).

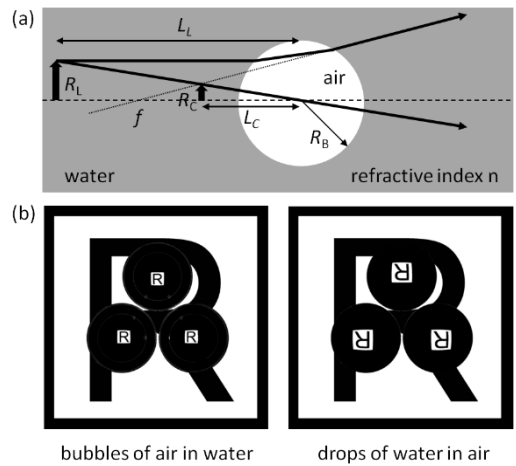

Figure 1: A bubble in water creates virtual image of an object with the same orientation. (a) Image construction of an air bubble in water (scheme not drawn to scale). (b) Images of the letter "R" on a square background taken through three bubbles of air in water (left) or three drops of water in air (right) .

The focal length $f$ of a thick lens is given by

$$\frac{1}{f} = \frac{n-n'}{n'}\left(\frac{1}{R_1} - \frac{1}{R_2}\right) + \frac{(n-n')^2}{n'n}\frac{d}{R_1 R_2}, \tag{1}$$

where $R_1$ and $R_2$ are the radii of curvature of either side of the lens and $d$ is the thickness of the lens. $n'$ is the refractive index of the lens, while $n$ is the refractive index of the medium surrounding the lens. In the case of a spherical bubble $R_1 = -R_2 = R_B$, $d = 2R_B$ and $n' \approx 1$.   Equ.(1) therefore simplifies to

$$\frac{1}{f} = 2(n-1)\frac{1}{R_B}. \qquad (2)$$

This means, for example, that a bubble with radius $R_B$ = 100 μm has a focal length of $f$ = 150 μm for n=1.33. Hence, the virtual image is generated very close to the bubble, which is practical for imaging purposes.

The relationship between the distance $L_L$ (distance between the object and the centre of the bubble) and $L_C$ (distance between the virtual image and the centre of the bubble) is related to the focal length $f$ of the bubble by simple optics

$$\frac{1}{f} = \frac{1}{L_C} - \frac{1}{L_L}. \qquad (3)$$

Moreover, we can relate those distances to the object and image size via simple geometry

$$\frac{L_L}{L_C} = \frac{R_L}{R_C}. \qquad (4)$$

Equating Equ.s (2) and (3), and using Equ. (4) one finds the relationship

$$R_B = 2(n-1)L_L\frac{R_C}{R_L - R_C}. \qquad (5)$$

Hence, knowing $L_L$ and $R_L$, and measuring $R_C$, one can obtain $R_B$. One notices that if $R_L$ is much larger than $R_C$ (in our experiments $R_L \approx 1000\ R_c$), the relationship between $R_B$ and $R_C$ can be approximated by

$$R_B = 2(n-1)\frac{L_L}{R_L}R_C. \qquad (6)$$

We use this simple idea to show how one can characterise in an automated and reliable way either a monodisperse population of bubbles (Section 0) or a polydisperse one (Section 4).

# 3  Imaging of small bubbles

We generate small bubbles by pushing a dishwashing solution (10% "Fairy" in tap water) repeatedly back and forth through a constriction of 1.5 mm diameter and 1 cm length by two syringes [9]. However, the technique used to generate the bubbles and the agents used to stabilise the bubbles have no influence on the optical measurement.

As sketched in Figure 2 the generated bubbles are spread on the surface of a petri-dish filled with the same solution, which is then covered by a glass plate in a way that the bubbles form a mono-layer. The petri-dish containing the bubble layer is then put onto a glass table, which is positioned above a circular light source. The circular light source is fabricated by cutting circles of different radii $R_L$ out of a black sheet which is then put onto a diffusive light sheet (Rex-Leuchtplatten). We use $R_L$ = 15 mm, 30 mm, 45 mm and 47 mm. The distance $L_L$ between the light sources and the bubble layer is 17 cm. These different dimensions should be chosen with respect to the specific application – as shall be clearer later on.

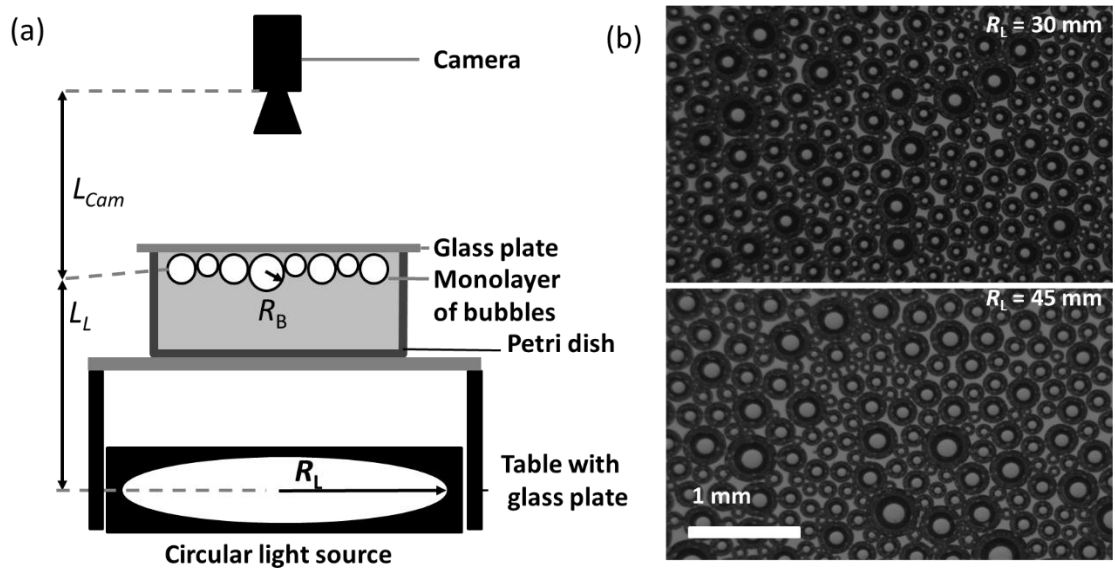

Figure 2: Set-up used to demonstrate the automated bubble size measurement: a circular light source is imaged through a monolayer of bubbles (set-up not drawn to scale). (b) Example of two images obtained for two different radii $R_L$ of the circular light source.

A digital camera (Ueye-1490) is positioned above the bubble layer such that the image of the circular light source can be seen through the bubbles. The focal length and the aperture of the objective should be chosen such that the focal depth of the image includes the bubbles and the images of the light source, i.e. that both are simultaneously in focus. The required focal depth is typically of a few hundred micrometres following the calculations given in Section 2. Here we use a Nikkon AF Micro objective (f = 60 mm) with two 40 mm rings for additional magnification. Two example images for light sources of different radii ($R_L$ = 30 mm and 45 mm) are shown in Figure 2 where the white circles, which are the image of the light source, can be clearly distinguished in the centres of the bubbles. We shall call the radii of the circles $R_C$. As can be seen in these images, multiple refraction/reflection creates a complex dark grey corona around the bubbles [7], which generally makes the application of other automated techniques (like Watershed algorithms) prone to errors. Due to these reflections, size and distance of the light source should be chosen in a manner that the boundary of the image of the light source does not penetrate into this corona.

The obtained images are then treated with an image analysis software. In our case we use ImageJ, freely available at http://imagej.nih.gov/ij/. As shown in Figure 3a, the images are thresholded and inverted. The software is then used to automatically find the circular objects in the images – since these correspond to the image of the light source. This can be done with high accuracy by limiting the range of areas and of circularity of the object to be accepted. An example of automatically detected objects is shown Figure 3a-4. After setting the proper scale of the images, the image analysis tool can then determine the radius $R_C$ of each image of the light source.

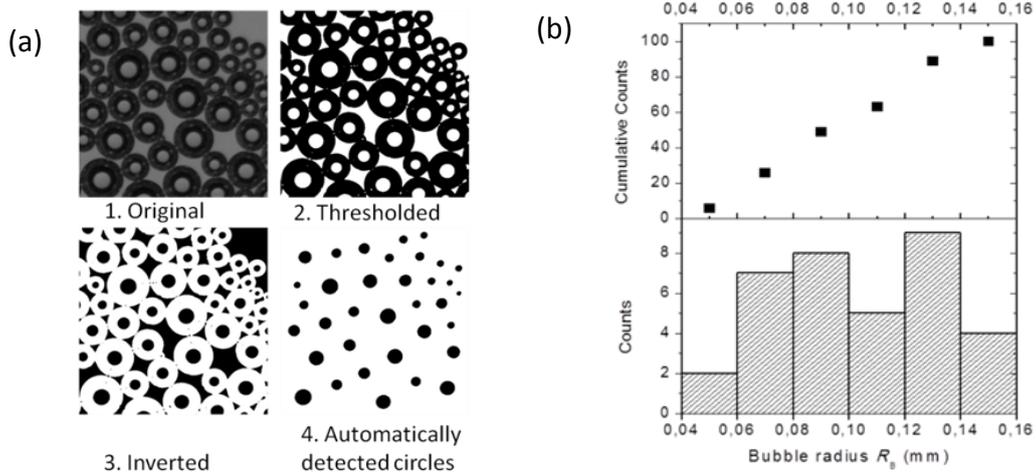

Figure 3: Illustration of the image treatment and typical result for $R_L$ = 45 mm. (a) The original image is thresholded and inverted before detecting in an automated manner the circular images of the light source. Using the calibration of the set-up with Equ. (5) the bubble radius $R_B$ can be automatically calculated. (b) Bubble size distribution obtained for (a).

In order to verify that Equ. (5) describes appropriately the experimental data, we use ImageJ to measure "by hand" the radius $R_B$ of a bubble along with the corresponding radius $R_C$ for a wide range of bubble sizes and for different light source dimensions $R_L$. The results are shown in Figure 4a together with the corresponding Equ. (5) for four different radii $R_L$ of light sources.

Figure 4a shows the data in a linear plot, showing that Equ. (5) corresponds to the experimental results only for a certain range of small bubble sizes. This range increases with the size $R_L$ of the light source. The same data is shown in a logarithmic plot to show the excellent agreement between theory and experiment for sufficiently small bubbles.

The reason why we observe this deviation is due to gravity: buoyancy forces press the bubbles against the top glass plate. As long as bubbles are sufficiently small, this has a negligible effect. However, when buoyancy forces become non-negligible, the top of the bubble is deformed into a flat zone of radius $r$, as shown in the inset of Figure 4b. One can show [10] that the overall bubble shape can be reasonably well approximated by a "truncated sphere" as long as the bubble radius remains much smaller than the capillary length

$$R_B < l_c = \sqrt{\frac{\gamma}{\Delta\rho g}}, \tag{7}$$

where $\gamma$ is the gas/liquid interfacial tension and $\Delta\rho$ and $g$ the density difference and gravity, respectively. In our experiments $\gamma = 34 \times 10^{-3}$ N/m, hence $l_c \approx 1.8$ mm. In equilibrium, the buoyancy force of the bubble has to be in equilibrium with the force exerted by the flat zone, which is equal to the area of the zone $\pi r^2$ times the pressure difference between the bubble and the liquid, which is the Laplace pressure $2\gamma/R_B$. One therefore obtains

$$V_B \Delta\rho g = \frac{4}{3}\pi R_B^3 \Delta\rho g = \pi r^2 \frac{2\gamma}{R_B}, \tag{8}$$

leading to

$$r = \sqrt{\frac{2}{3}\frac{R_B^2}{l_c}}. \tag{9}$$

Since the truncated part of the sphere remains small in comparison to the overall volume, we can assume that the radius of the truncated sphere is the same as that of the undisturbed bubble. However, the truncation plays an important role for the lensing properties. As soon as the light rays pass through this zone the imaging deviates from that of a perfect spherical lens. If we approximate that this happens when the image size is roughly of the order of the flat zone, we can replace $r$ in Equ. (9) by $R_C$ in order to obtain an approximation beyond which $R_B(R_C)$ one expects the data to deviate from the simple theory given in Equ. (5). One obtains

$$R_B = \left(\frac{2}{3}\right)^{1/4}\sqrt{R_C l_c}. \tag{10}$$

This is shown by the gray, dashed line in Figure 4. Above this line gravitational deformation of the bubbles should lead to less strong lensing effects, the theory of the spherical bubbles being no longer suitable, which is indeed observed in the deviation from Equ. (5) above this line.

This means that in order to use Equ. (5) for a maximum range of bubble sizes, one needs to work with a sufficiently large light source. However, even if the data deviates from Equ. (5), a simple, non-linear fit of the data can be used to calibrate the set-up as long as the deformations of the bubbles remain small.

The established $R_B(R_C)$ - relationship can then be used to translate the automatically detected images of size $R_C$ of the light source into the corresponding bubble sizes. An example of the obtained bubble size distribution is shown in Figure 3b for the image of Figure 3a. In order to maintain clarity of the images, only a few bubbles were used in the example. However, with high resolution cameras, up to a few thousand bubbles can be treated reliably in one image. Moreover, since the treatment can be fully automated once the setup is calibrated, it can be used to treat entire videos and therefore establish the variation of bubble size distributions over time.

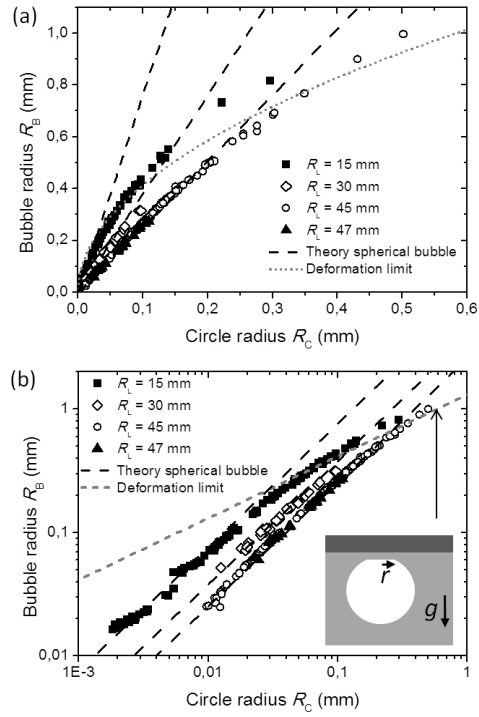

**Figure 4: (a) Relationship between the bubble radius $R_B$ and the radius $R_C$ of the image of the circular light source for different radii RL of the light source. The dashed, black lines correspond to Equ. (5) without any fitting parameters. It can be seen that this equation fits very well the data until a gravity-driven deviation occurs above the gray, dashed line which is given by Equ. (5). (b) log-log presentation of (a).**

## 4 Large and polydisperse bubbles

When dealing with strongly polydisperse bubble size distributions, containing simultaneously large and small bubbles, the automated treatment used as explained in Section 0 has two main difficulties. The first one is that large bubbles are too strongly deformed by gravity so that they cannot be approximated any more by a capped sphere, as done in Section 3. The second involves the difficulty to catch properly the small bubbles. When the bubbles are deposited on the solution it is not trivial to make a true monolayer since the small bubbles tend to "snuggle" underneath larger ones. Moreover, it becomes difficult to distinguish what is a small bubble and what is a gap between bubbles, even when tuning the circularity criteria as explained before.

To overcome these different difficulties it is more appropriate to squeeze the bubbles between two narrowly spaced glass plates of known spacing $2h$, as sketched in Figure 5a. To obtain this layer, a small droplet of the foaming solution is placed on a glass plate. The bubbles are then deposited onto that droplet and an upper glass plate is disposed which ensures a good spreading of the bubbles into a monolayer. Both plates are separated by plate spacers of well-defined thickness $2h$ which ensure that the spacing is homogenous and known with high precision. In our case we use microscope cover slides for this purpose with $2h = 150$ μm.

The circular light source used here is $R_L = 45$ mm and the distance between the light source and the bubble layer is $L_L = 16$ cm.

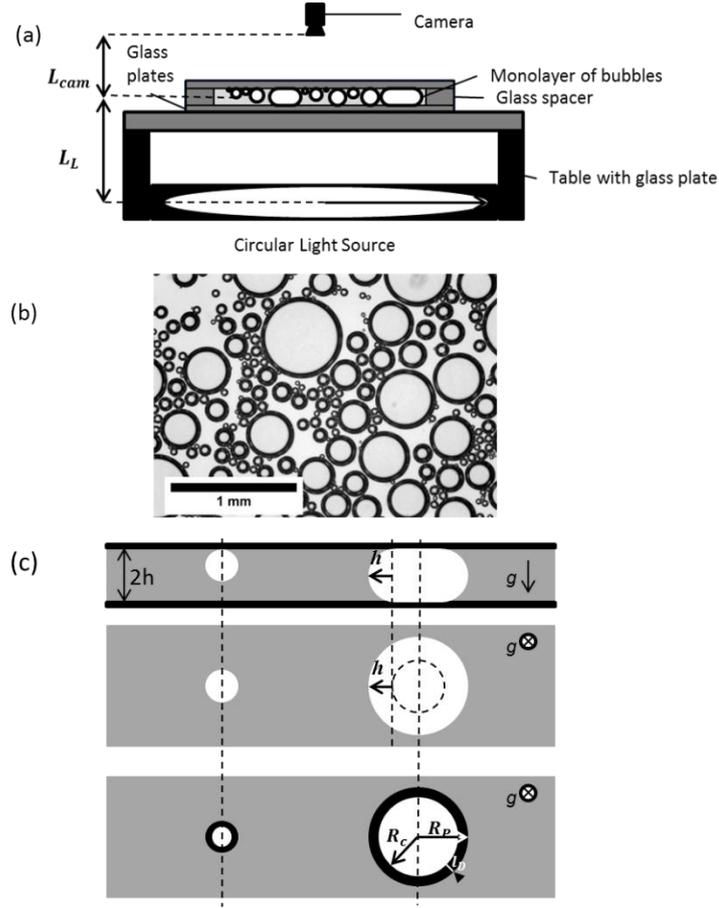

**Figure 5: (a)** Set-up used to demonstrate the automated bubble size measurement of wide size distribution: a circular light source is imaged through a monolayer of bubbles sandwiched between two glass plates (set-up not drawn to scale). **(b)** Obtained image with a circular light source of $R_L$ = 45 mm. **(c)** Sketch of bubbles when their diameter is smaller (*Left*) or larger (*Right*) than the height of the gap between glass plates with accompanying definition of the variables used in the text. The bottom row corresponds to the images seen in the photographs containing the black ring around the bubble.

As in Section 0, images are treated using ImageJ. However, this time one needs to be careful because bubbles are no longer spherical when their diameter $2R_B$ is larger than the separation $2h$ of the glass plates. As shown in top row of Figure 5c, the bubbles are then flattened into a pancake shape of overall radius $R_P$ which is composed of a central cylinder of height $2h$ and radius $R_P-h$, and a boundary whose cross-section is a semi-circle with radius of curvature $h$. The volume $V_P$ of such a pancake is given by

$$V_p = 2\pi h(R_P - h)^2 + \pi^2 h^2(R_P - h) + \frac{4}{3}\pi h^3, \tag{11}$$

When bubbles are only slightly deformed, their shapes are not true pancakes [10]. In order to estimate the error one makes by using Equ. (11) we performed Surface Evolver simulations [8] (http://www.susqu.edu/brakke/evolver/evolver.html) to obtain the true bubble shape. Figure 6 shows images of the resulting bubble shapes with increasing bubble volume, and Figure 6b compares the results of the simulations with the approximation of the pancake shape. As one can see, the error one makes in using the pancake approximation is negligibly small for our purposes.

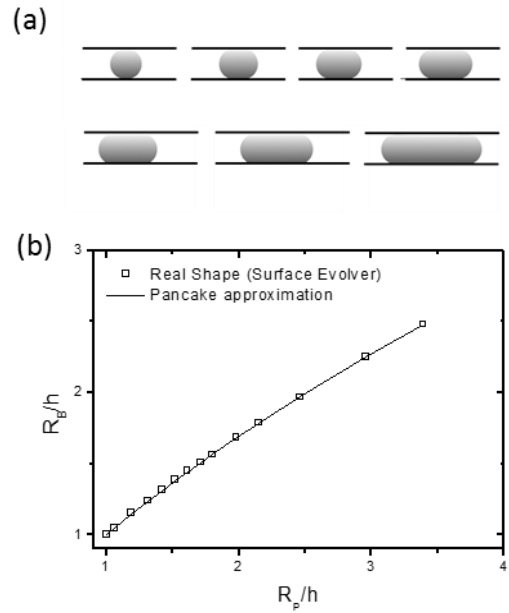

**Figure 6: (a) Shapes of bubbles squeezed between glass plates obtained using Surface Evolver Simulations. (b) Radius of the unsqueezed bubble as a function of the radius of the squeezed bubble $R_P$, comparing the pancake approximation of Equ. (12) and the simulations shown in (a).**

In order to obtain the radius of the unsqueezed bubble one must first determine the relationship between the in-plane radius $R_P$ and the circle radius $R_c$, and then calculate the true bubble radius $R_B$ from the pancake approximation given in Equ. (11). Using $R_B = \left(\frac{3}{4\pi} V_p\right)^{1/3}$ one obtains

$$R_B = \left(\left(\frac{3}{4\pi}\right)\left(2\pi h(R_P - h)^2 + \pi^2 h^2(R_P - h) + \frac{4}{3}\pi h^3\right)\right)^{1/3} \qquad (12)$$

This equation can be used to calculate the bubble radius $R_B$ from the value of the circle radius $R_C$ via

$$R_P = R_C + \delta. \tag{13}$$

Knowing that the boundary of the pancake images the light source in the same way as spherical bubbles, we can use Equ. (6) to relate h and δ

$$h = 2(n'-1)\frac{L_L}{R_L}(h-\delta), \tag{14}$$

leading to

$$h - \delta = \alpha h, \tag{15}$$

with

$$\alpha = \frac{1}{2(n'-1)}\frac{R_L}{L_L}. \tag{16}$$

Combining these expressions with Equ. (12) one obtains the relationship between $R_B$ and $R_C$

$$R_B = \left[\frac{3}{2}h(R_C - \alpha h)^2 + \frac{3\pi}{4}h^2(R_C - \alpha h) + h^3\right]^{1/3}. \tag{17}$$

As in Section 0, we perform a series of measurements "by hand" where we measure for a wide range of bubble sizes the bubbles radius and the corresponding circle radius $R_C$ in the images using ImageJ. For bubbles smaller than the plate spacing we apply the procedure of Section 0, while for bubble larger than the plate spacing we apply the pancake approach presented in this section. The results for both regimes are shown in Figure 7a. The transition between the two regimes can be clearly observed at $R_B = h$. For $R_B < h$ the solid line corresponds to Equ. (6), while for $R_B > h$ the dashed line is given by Equ. (17) (in conjunction with Equ. (16)). One can see that in both cases the experimental data is very well described by the theory without any fitting parameters. The theoretical predictions can therefore be used to automate the image treatment: the computer program finds the white circles and applies the two different calibrations for the spherical and for the pancake bubbles independently. An example of the result of the automated treatment of Figure 5b using the approach described above is plotted in Figure 7b.

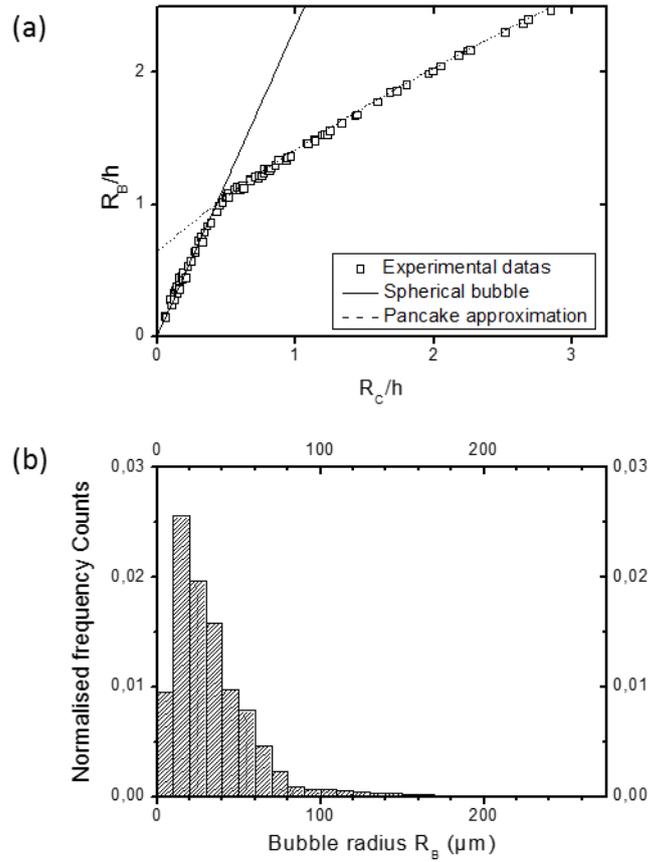

**Figure 7: (a) Result of the calibration. The straight line corresponds to the theory for small bubbles of Equ. (5), while the dashed line corresponds to the pancake approximation as expressed in Equ. (17) (b) Result of automated analysis of Figure 5b using this measurement technique**

# 5 Conclusion and discussion

We have shown here a simple approach to measure automatically the size distributions of bubbles making use of the imaging properties of bubbles with a light source of well-defined geometry. We considered two different limits. In the first case, bubbles are floating underneath a horizontal surface. As long as the bubbles are small enough (following the conditions in Equ.s (7) and (10)) so that the gravitational deformation remains negligible, the data which relates the bubble size to the size of the imaged object through the bubble is well described by Equ. (5). This provides a particularly wide range of bubble sizes when using a large light source, as can be seen from Equ. (10). Once bubble sizes become too large to neglect gravity it is more appropriate to image the bubbles between two narrowly spaced plates. This is described in Section 4, where we also provide optical and geometrical calculations which allow to relate the detected image of the light source to the bubbles size with high precision and without any fitting parameter (Equ. (17)). In polydisperse foams, bubbles can be larger and smaller than the plate spacing, which is why one commonly needs to combine the regimes of spherical and pancake bubbles. However, the transition point between both regimes is clearly detectible on the data (as seen in Figure 7a) and also comes out directly from calculations by considering the cross-over of the theoretical models of each limit. The transition can therefore be taken into account in a fully automated manner.

Despite their power, both techniques have certain limits. One problem arises from the fact that bubbles need to form a true monolayer. In highly polydisperse systems, small bubbles may "snuggle" underneath larger ones, even when squeezed between narrowly spaced plates, making our imaging technique invalid. Hence, the technique is best applied to bubble size distributions with limited polydispersity. Moreover, when working with a closely-packed monolayers of bubbles, stability of the bubbles against coalescence is an important ingredient to reliable measurement of distributions. If this stability cannot be assured, for example through the addition of a surfactant, one may want to resort to an approach in which bubbles float freely at low density. The imaging of freely floating bubbles will also remove the problems of bubble deformation under gravity in contact with a plate.

Despite these limitations, we hope to have shown that using the lensing properties of bubbles in order to determine automatically their size distributions can be a powerful tool. We have chosen here the simples object geometry – a circular light source – but any other geometry is possible in order to facilitate automatic detection of the images.